\documentclass[aps, prl, reprint,10pt,longbibliography,superscriptaddress]{revtex4-2}

\usepackage[bookmarks=true,colorlinks,citecolor=blue,urlcolor=blue]{hyperref}
\usepackage[usenames,dvipsnames]{xcolor}

\usepackage{amssymb}
\usepackage{graphicx}
\usepackage{mathtools}% loads amsmath
\usepackage{braket}
\usepackage{blindtext}
\usepackage[normalem]{ulem}

\usepackage[toc,page,header]{appendix}
\usepackage{minitoc}
\usepackage{tocloft}
\newcommand*{\Scale}[2][4]{\scalebox{#1}{\ensuremath{#2}}}%

\definecolor{pink2}{rgb}{0.858, 0.188, 0.478}

\newcommand{\Tr}[1]{\mathrm{Tr}#1}

\newcommand{\eq}[1]{Eq.\thinspace(\ref{#1})}

\newcommand{\fc}[1]{({#1})}
\newcommand{\figc}[2]{Fig.\thinspace{}\ref{#1}\thinspace{}\fc{#2}}

\begin{document}

\doparttoc % Tell to minitoc to generate a toc for the parts
\faketableofcontents % Run a fake tableofcontents command for the partocs

\part{} % Start the document part
%\parttoc % Insert the document TOC

\title{Coupled Hydrodynamics in Dipole-Conserving Quantum Systems}

\newcommand{\TUM}{\affiliation{Department of Physics, Technical University of Munich, 85748 Garching, Germany}}
\newcommand{\MCQST}{\affiliation{Munich Center for Quantum Science and Technology (MCQST), Schellingstr. 4, 80799 M{\"u}nchen, Germany}}

\author{Ansgar G. Burchards}  \TUM \MCQST
\author{Johannes Feldmeier} \TUM \MCQST
\author{Alexander Schuckert} \TUM \MCQST
\author{Michael Knap} \TUM \MCQST

\begin{abstract}
We investigate the coupled dynamics of charge and energy in interacting lattice models with dipole conservation. We formulate a generic hydrodynamic theory for this combination of fractonic constraints and numerically verify its applicability to the late-time dynamics of a specific bosonic quantum system by developing a microscopic non-equilibrium quantum field theory. Employing a self-consistent $1/N$ approximation in the number of field components, we extract all entries of a generalized diffusion matrix and determine their dependence on microscopic model parameters. We discuss the relation of our results to experiments in ultracold atom quantum simulators.
\end{abstract}

\maketitle

\section{Introduction}

Understanding the non-equilibrium dynamics of interacting quantum many-body systems and their general governing principles is a fundamental challenge. Conventional wisdom holds that global conserved quantities in such systems lead to diffusive transport at late times~\cite{chaikin_principles_2000, Mukerjee06, lux_hydrodynamic_2014, bohrdt_scrambling_2017, leviatan_quantum_2017}. However, recently a variety of systems with unconventional transport properties have been identified, including superdiffusion  in integrable systems~\cite{ljubotina_kardar-parisi-zhang_2019, bulchandani_2021, wei_quantum_2021} and in the presence of long-range interactions~\cite{schuckert_non-local_2020, joshi_observing_2021, zu2021emergent}, as well as subdiffusive transport in disordered many-body systems~\cite{agarwal2015_subdiff,znidaric2016_subdiff, vosk2015_mbl, Potter2015, Agarwal2017}.
In this context, novel quantum systems with constraints have recently attracted much attention. In particular, fracton models~\cite{nandkishore_fractons_2019,pretko_fracton_2020,chamon_quantum_2005,haah_local_2011,yoshida2013_fractal,Vijay15,Vijay16} that conserve a global $U(1)$ charge and its associated dipole moment~\cite{pretko_subdimensional_2017,pretko_fracton_2018,Pretko17b,
williamson2019_fractonic} display a variety of exotic non-equilibrium phenomena: At sufficiently sparse filling, ergodicity is broken due to a `strong fragmentation' of the many-body Hilbert space into disjoint subsectors~\cite{sala_ergodicity-breaking_2020,khemani_localization_2020,rakovszky_statistical_2020,Scherg2021aa, kohlert_experimental_2021}, limiting the growth of entanglement and inducing slow spreading of operators~\cite{feldmeier2021_operator}. Even in more generic ergodic situations, the conservation of the dipole moment qualitatively changes the dynamics, giving rise to a new universality class of subdiffusive hydrodynamics~\cite{guardado-sanchez_subdiffusion_2020,gromov_fracton_2020,feldmeier_anomalous_2020,morningstar_kinetically_2020}. Experimentally, these aspects of constrained dynamics can be probed in ultracold atom quantum simulators where a strong, linearly tilted potential enforces the conservation of the dipole moment~\cite{guardado-sanchez_subdiffusion_2020,Scherg2021aa, kohlert_experimental_2021}.

\begin{figure}
\includegraphics[width=.48\textwidth]{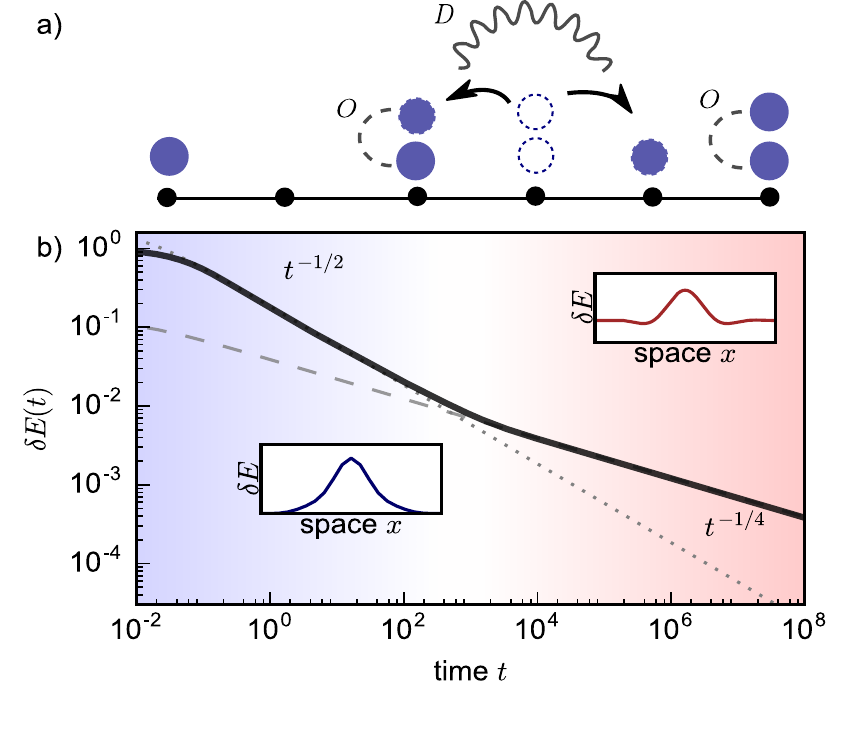}
\caption{\textbf{Coupled hydrodynamics in fractonic quantum matter.} a) Illustration of the dipole-conserving dynamics within our microscopic field-theory. Bosons (blue circles) situated on lattice sites need to coordinate with other bosons in order to move. This coordinated hopping, as well as the local boson-boson repulsion, are mediated by exchange field propagators $D$ and $O$, respectively. b) Decay of an energy excitation including a small charge-density excitation obtained from dipole-conserving hydrodynamics. 
A diffusive regime with dynamical exponent $z=2$ emerges at intermediate times and crosses over to a subdiffusive regime with $z=4$ at late times. The latter arises from the charge contribution to the excitation and is a signature of dipole conservation. Insets: Spatial profiles of the excitation at intermediate (blue) and late (red) times.}
\label{fig:coverIllustration}
\end{figure}

Recent theoretical works have derived universal transport properties in charge- and dipole-conserving systems starting directly from effective hydrodynamic equations, with~\cite{glorioso_breakdown_2021,osborne2021infinite,grosvenor2021hydrodynamics,glorioso_toappear} and without~\cite{gromov_fracton_2020,feldmeier_anomalous_2020} momentum-conservation.
A verification of these results has been mostly limited to random unitary circuit dynamics in the absence of energy conservation~\cite{feldmeier_anomalous_2020,morningstar_kinetically_2020,iaconis_multipole_2021}; see however~\cite{glorioso_breakdown_2021}. Therefore, it remains an open challenge to investigate the coupled hydrodynamics of charge and energy emerging in microscopic Hamiltonian lattice systems with energy and dipole conservation.

In this work, we provide such an analysis by studying the non-equilibrium dynamics of a one-dimensional system of lattice bosons with charge-, dipole- and energy-conservation. We derive a general set of equations within the framework of linear fluctuating hydrodynamics that take into account couplings between charge- and energy-excitations in the presence of dipole conservation; Sec.~\ref{sec:hydrodynamics}. This results in a subdiffusive mode of mixed charge-energy excitations as well as a diffusive pure energy mode; see \figc{fig:coverIllustration}{b} for the former. We compare this effective hydrodynamic picture with the quantum time evolution of a specific bosonic lattice model. To this end, we develop a non-equilibrium quantum field theory starting from the microscopic Hamiltonian in Sec.~\ref{sec:methods}. Using the 2-particle-irreducible (2PI) effective action approach, we derive equations of motion for Green's functions that describe the relaxation of densities associated with conserved quantities. Solving these equations numerically, we verify the applicability of our mode-coupled hydrodynamics in Sec.~\ref{sec:charge_and_energy_dynamics} and  extract all components of a generalized diffusion matrix. An outlook and a discussion of how our results are related to quantum simulation experiments are provided in Sec.~\ref{sec:conclusion}. Technical details are deferred to the appendix.

\section{Mode-coupled Hydrodynamics}
\label{sec:hydrodynamics}
We consider a one-dimensional quantum system whose time evolution is generated by a microscopic lattice Hamiltonian $\hat{H}$ that conserves a global U(1) charge $\hat{Q}$ as well as its associated dipole moment $\hat{P}$. Expressed in terms of microscopic charge and energy densities $\hat{n}_x$ and $\hat{h}_x$, the conserved quantities read
\begin{equation} \label{eq:1.11}
\hat{H} = \sum_x \hat{h}_x, \quad \hat{Q} = \sum_x \hat{n}_x, \quad \hat{P} = \sum_x x\, \hat{n}_x.
\end{equation}
Our goal is to formulate a long-wavelength description of transport for the set of conserved quantities \eqref{eq:1.11} within linear fluctuating hydrodynamics. In this approach, hydrodynamic equations for the conserved macroscopic densities are combined with noise terms accounting for fluctuations that are generated by the underlying microscopic degrees of freedom.

Let us first consider the continuity equation of a single conserved quantity, such as energy
\begin{equation} \label{eq:1.12}
\partial_{t} e(x,t) = - \partial_x j^{e}(x,t),
\end{equation}
where $e(x,t)$ and $j^e(x,t)$ are the coarse-grained energy and current densities. Taking the long-wavelength limit, a gradient expansion of the current can be performed. The first order term leads to Fick's law $j^e(x,t) = -D_e \partial_x e(x,t)$. Including current fluctuations $\xi^e(x,t)$, we obtain the fluctuating hydrodynamic equation for a single diffusive mode,
\begin{equation} \label{eq:1.13}
\partial_t e(x,t) - D_e \partial_x^2 e(x,t)= \partial_x \xi^e(x,t) \ , 
\end{equation}
where space and time are related by the dynamical exponent $z=2$ and $\braket{\xi^e(x,t)\xi^e(x^\prime,t^\prime)} = B_e^2 \delta(x-x^\prime)\delta(t-t^\prime)$ describes uncorrelated Gaussian white-noise with noise strength $B_e$.

The time evolution of the coarse-grained charge density $n(x,t)$ is drastically modified by dipole conservation. The fundamental dynamical objects in the system are \textit{dipoles}, a fact we can incorporate by introducing a coarse-grained dipole current $j^d(x,t)$. Importantly, as opposed to a more conventional charge current, $j^d(x,t)$ is \textit{even} under spatial inversion $\hat{I}$, i.e. $\hat{I} j^d(x,t) \hat{I} =  j^d(-x,t)$. Thus, the current transforms under inversion in the same way as the underlying charge density $n(x,t)$, and the corresponding continuity equation assumes the form~\cite{gromov_fracton_2020, feldmeier_anomalous_2020}
\begin{equation} \label{eq:1.14}
\partial_t n(x,t) = \partial_x^2 j^d(x,t).
\end{equation}
In the absence of energy conservation, the generalized Fick's law $j^d(x,t) = -D_n \partial_x^2 n(x,t)$ then preserves inversion invariance of the resulting hydrodynamic equation
\begin{equation} \label{eq:1.15}
\partial_t n(x,t) + D_n \partial_x^4 n(x,t) = \partial_x^2 \xi^d(x,t).
\end{equation}
Here, we included dipole-current fluctuations $\xi^d(x,t)$ with $\braket{\xi^d(x,t)\xi^d(x^\prime,t^\prime)} = B_n^2 \delta(x-x^\prime)\delta(t-t^\prime)$. \eq{eq:1.15} describes a \textit{subdiffusive} mode with dynamical exponent $z=4$ and has previously been observed numerically in unitary circuit models without energy conservation~\cite{feldmeier_anomalous_2020,moudgalya2021_spectral,iaconis_multipole_2021}. 

In systems exhibiting the full set of conserved quantities \eqref{eq:1.11}, we must find a description that combines \eq{eq:1.13} and \eq{eq:1.15} while taking into account possible couplings between the charge and energy densities. This amounts to including cross-terms in the derivative expansion of the currents $j^{d}$ and $j^{e}$:
\begin{equation} \label{eq:1.16}
\begin{split}
j^{d} =  -D_{nn} \partial_{x}^2 n - D_{ne} \partial_{x}^{2} e \\
j^{e} =  -D_{en} \partial_{x} n - D_{ee} \partial_{x} e.
\end{split}
\end{equation}
\eq{eq:1.16} includes the most relevant (i.e. fewest derivatives) terms compatible with inversion symmetry of the resulting hydrodynamic equations. A potential coupling between the microscopic current fluctuations vanishes: $\xi^e\xi^d$ is odd under inversion according to our previous considerations, such that the expectation value $\braket{\xi^e\xi^d}=0$ in an inversion invariant equilibrium ensemble. Combining \eq{eq:1.16} with the continuity equations (\ref{eq:1.12}, \ref{eq:1.14}) and switching to momentum space, our Ansatz becomes
\begin{equation} \label{eq:1.18}
\bigl[
\partial_t
 +
k^2
\underline{D}(k)
\bigr]
\begin{pmatrix}
 n \\
 e
\end{pmatrix} 
=
ik
\underline{B}(k)
\begin{pmatrix}
 \xi^{d} \\
 \xi^{e}
\end{pmatrix},
\end{equation}
where we have defined the matrices
\begin{equation} \label{eq:1.19}
\underline{D}(k) = 
\begin{pmatrix}
k^2 D_{nn} & k^2 D_{ne} \\
D_{en} & D_{ee}
\end{pmatrix}, \quad
\underline{B}(k) = 
\begin{pmatrix}
ik B_n & 0 \\
0 & B_e
\end{pmatrix}.
\end{equation}
%Comparing the relaxation dynamics of a particular microscopic theory with \eq{eq:1.20} thus provides three of the four couplings entering the diffusion matrix $\underline{D}(k)$ in \eq{eq:1.19}. 
As dipole conservation implies a diagonal matrix $\underline{B}(k)$, only two of the diffusion constants entering $\underline{D}(k)$ are independent. This can be seen by considering the fluctuation-dissipation relation associated with \eq{eq:1.18} (see also App.~\ref{app:fdrs}):
\begin{equation} \label{eq:1.21}
\underline{D}(k)\,\underline{C} + \underline{C}\,\underline{D}^T(k) = \underline{B}(k)\underline{B}^T(-k),
\end{equation}
where $\underline{C}$ is the matrix of static equilibrium correlations. The Onsager relations for kinetic coefficients require that $\underline{D}(k)\underline{C}$ is symmetric. Using \eq{eq:1.21} this is equivalent to \begin{equation} \label{eq:1.22}
C_{nn}D_{en}+C_{en}D_{ee} = 0,
\end{equation}
which further implies
\begin{equation} \label{eq:1.23}
C_{en}D_{nn} + C_{ee}D_{ne} = 0.
\end{equation}
When the equilibrium correlations are known (see App.~\ref{app:Infinite_T_corr}), 
Eqs.~(\ref{eq:1.22}, \ref{eq:1.23}) can be used to determine two of the diffusion constants entering \eq{eq:1.19}.

Solving the coupled hydrodynamic equations \eqref{eq:1.18} in the long-wavelength limit predicts the existence of two independent modes, a diffusive energy-only mode as well as a subdiffusive energy-charge mode. In particular, inhomogeneities of the initial state decay according to 
\begin{equation} \label{eq:1.20}
\begin{pmatrix}
n(k, t) \\ e(k, t)
\end{pmatrix}
= a\, e^{-D_{ee}k^2 t}
\begin{pmatrix}
0 \\ 1
\end{pmatrix}
+ b\, e^{-\tilde{D}_{nn}k^4 t}
\begin{pmatrix}
1 \\ -\frac{D_{en}}{D_{ee}}
\end{pmatrix},
\end{equation}
where $\tilde{D}_{nn} = D_{nn} - \frac{D_{ne} D_{en}}{ D_{ee}} $ is the renormalized subdiffusion constant governing the decay of the mixed energy-charge mode and the constants $a,b$ are fixed by the initial state.
In the following, we will study the dynamics of a specific, strongly interacting bosonic system using non-equilibrium quantum field theory. Within this approach we verify that \eq{eq:1.20} provides an accurate description of the quantum evolution emerging at late times. Furthermore, comparing the microscopic dynamics to \eq{eq:1.20} we can extract the two constants $D_{ee}$ and $\tilde{D}_{nn}$. Together with Eqs.~(\ref{eq:1.22}, \ref{eq:1.23}) this allows us to directly link all entries of the diffusion matrix \eqref{eq:1.19} to model parameters of the microscopic Hamiltonian.

\section{Model and Microscopic Field Theory}
\label{sec:methods}
We study the dipole-conserving lattice Hamiltonian
\begin{equation}
\label{eq:Hamiltonian}
H = \underbrace{J \Big( \sum_i \hat{\phi}_{i-1}^\dagger \hat{\phi}_{i}^{2} \hat{\phi}_{i+1}^\dagger +\text{h.c.} \Big)}_{\quad\Scale[1]{H_{\text{pair}}}}  + \underbrace{U \sum_i \hat{n}_{i} (\hat{n}_{i} - 1)}_{\quad\Scale[1]{H_{U}}} \, ,
\end{equation}
where $\hat{\phi}_{i}^{\dagger}/ \hat{\phi}_{i}$ denote the bosonic creation/annihilation operators and $\hat{n}_{i} = \hat{\phi}_{i}^{\dagger}\hat{\phi}_{i}$ the occupation number operator on site $i$. The first term describes a short-ranged and dipole-conserving hopping of bosons which can be interpreted as a kinetic term for particle-hole pairs. 
The second term captures local repulsion between bosons which competes with the hopping strength. The equilibrium phase diagram of such dipole-conserving Bose-Hubbard models has recently been discussed in Ref.~\cite{lake2022}. We consider fillings above unity, where hydrodynamic behaviour is not expected to be inhibited by effects such as strong Hilbert space fragmentation ~\cite{sala_ergodicity-breaking_2020, morningstar_kinetically_2020}. This regime is challenging to investigate by numerical methods such as exact diagonalization and matrix product states due to the unrestricted local Hilbert space dimension, the rapid build up of entanglement, as well as the large systems and late times required to reach the hydrodynamic regime. Our microscopic, non-equilibrium field theoretic approach overcomes these limitations at the expense of approximating interaction effects. 

\subsection{Non-Equilibrium Field Theory }
In the following, we will derive self-consistent equations of motion for the correlations that characterize transport of charge and energy. Here, we outline our approach and defer further technical details to App. \ref{app:eoms}.
\begin{figure*}[!ht]
\includegraphics[width=\textwidth]{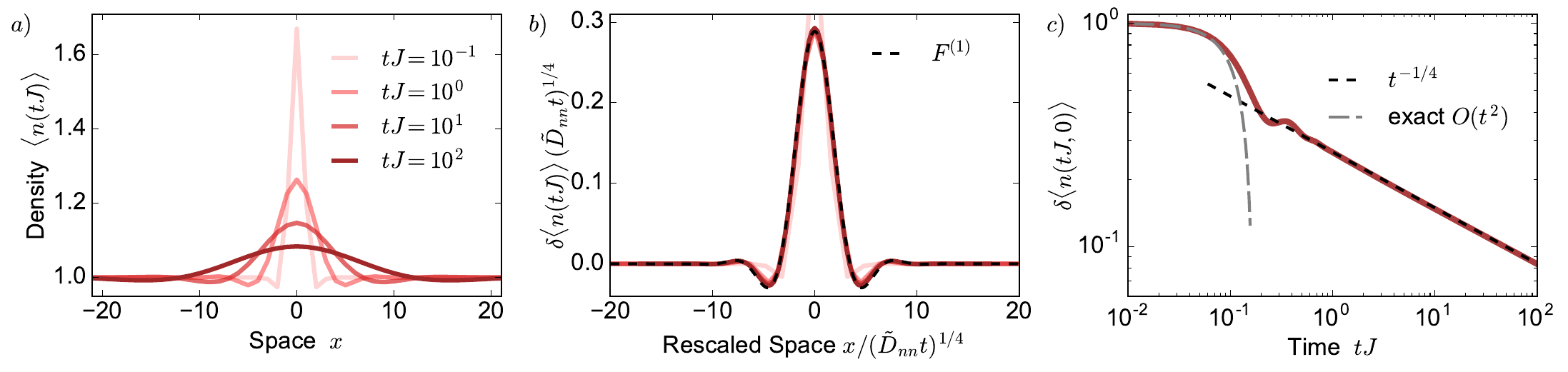}% Here is how to import EPS art
\caption{\label{fig:ScalingN} \textbf{Evolution of a non-equilibrium initial state with a sharply peaked charge excitation.} a) Charge profile at various times. b) The rescaled charge profile collapses to the scaling function $F^{(1)}$. c) The local excitation decays algebraically with exponent close to $1/4$. The short time dynamics agree with exact results obtained up to quadratic order. }
\end{figure*}
We start from the microscopic action 
\begin{equation} \label{eq:1.1}
    S[\phi] = \int_{\mathcal{C}} dt \,  \overline{\phi}_{i} i \partial_{t_1} \phi_{i} - H(\overline{\phi}, \phi) \ ,
\end{equation}
where $\mathcal{C}$ indicates integration along the Schwinger-Keldysh contour and $H(\overline{\phi}, \phi)$ denotes the normal-ordered Hamiltonian with operators $\hat{\phi}_i$ ($\hat{\phi}_i^\dagger$) replaced by complex fields $\phi_i$ ($\bar{\phi}_i$). We remove all quartic terms by introducing a complex decoupling field $\chi$, which removes the quartic term $H_\mathrm{pair}$, as well as a real field $\Delta$ removing the term $H_U$. This yields the modified action

\begin{equation} \label{eq:1.2}
\begin{split}
S[\phi, \chi, \Delta] & =  \int_{\mathcal{C}} dt \, \overline{\phi}_i i \partial_t \phi_i + \overline{\chi}_{i} V^{-1}_{i j }\chi_{j} +  \frac{1}{2}\Delta_{i} U^{-1} \Delta_{i} \\ & - \sqrt{2} \big (\overline{\chi}_{i} \overline{\phi}_{i} \phi_{i+1}  +  \overline{\phi}_{j+1} \phi_{j} \chi_{j} + \overline{\phi}_{i} \phi_{i} \Delta_{i} \big )\ ,
\end{split} 
\end{equation}
where $V_{ij} = \frac{1}{2}(\delta_{i j + 1} + \delta_{i j - 1})$. The fundamental dynamical objects for which we derive equations of motion are the connected field correlators
\begin{equation} \label{eq:1.3}
\begin{split}
G_{ij}(t_1, t_2)  = & \Big \langle T_{\mathcal{C}} \,  \overline{\phi}_{i}(t_1) \phi_{j}(t_2) \Big \rangle =  \begin{gathered}
\includegraphics[width=2cm]{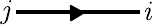}
\end{gathered}\\
D_{ij}(t_1, t_2) = & \Big \langle T_{\mathcal{C}} \, \overline{\chi}_{i}(t_1) \chi_{j}(t_2) \Big \rangle =  \begin{gathered}
\includegraphics[width=2cm]{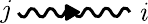}
\end{gathered} \\
    O_{ij}(t_1, t_2)  = &\Big\langle T_{\mathcal{C}} \, \Delta_{i}(t_1) \Delta_{j}(t_2) \Big \rangle =  \begin{gathered}
\includegraphics[width=2cm]{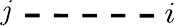}
\end{gathered} \ ,
\end{split} 
\end{equation}
where $T_{\mathcal{C}}$ describes time ordering on the Keldysh contour. The correlator $G_{ij}$ characterizes the dynamics of bosons and energy (see App. \ref{app:Energy_density}), while $D_{ij}$ and $O_{ij}$ contain information related to the 'dipole-dipole' and density-density correlators $\langle T_{\mathcal{C}} \, \overline{\phi}_{i}(t_1) \phi_{i+1}(t_1) \overline{\phi}_{j}(t_2) \phi_{j-1}(t_2) \rangle$ and $\langle T_{\mathcal{C}} \, \hat{n}_{i}(t_1) \hat{n}_{j}(t_2) \rangle$ respectively.

Our approach is based on deriving equations of motion for these correlators. 
This is achieved via the 2PI effective action $\Gamma^{\text{2PI}}$, which may be thought of as a quantum analogue of the classical action $S$ \cite{coleman_quantum_2019}:
Classical actions generate equations of motion via a stationarity condition with respect to a classical systems' path through phase space.
Similarly, the 2PI effective action yields equations of motion for Green's functions from a set of stationarity conditions with respect to these Green's functions \cite{berges_introduction_2004}. More specifically,
\begin{equation}
\label{eq:dGamma/DG_mainText}
    \frac{\delta\Gamma^{\text{2PI}}[G, D, O]}{\delta K_{ij}(t_1, t_2)} = 0,  
\end{equation}
for $K = G, D, O$.  Here, we implicitly assume symmetry unbroken initial states with $\langle \hat{\phi}_{i}(t)\rangle=0$, which also implies $\langle \chi \rangle = 0$, and further omit the dependence of  $\Gamma^{\text{2PI}}$ on $\langle \Delta \rangle$ through the exact relation $\langle \Delta_{i}(t) \rangle = \sqrt{2}U G_{i}(t,t)$. The resulting equations of motion for the full set of correlators $G, D, O$ can be solved by using a predictor-corrector method to propagate a square in the $t_1 - t_2$ plane. All propagator components needed for the forward propagation are known after bringing the equations of motion into explicitly causal form~\cite{schuckert_non-equilibrium_2018} (see App. \ref{app:eoms}). 
$\Gamma^{\text{2PI}}$ can be decomposed into up to 1-loop parts, containing the mean-field contributions, plus higher order parts $\Gamma^{\text{2PI}} = \Gamma^{(\text{1loop})} + \Gamma_2$. While the 1-loop parts possess a closed-form expression, the higher order parts $\Gamma_{2}$ are given by the sum over all vacuum 2PI diagrams and include scattering contributions necessary to describe the approach to thermal equilibrium~\cite{berges_thermalization_2001, aarts_far--equilibrium_2002}. Approximating this expression by any subset of such diagrams constitutes a conserving approximation which respects the underlying microscopic symmetries and therefore global conservation laws~\cite{baym_self-consistent_1962}. This is an important feature required to accurately describe late-time hydrodynamical behavior. 

We approximate $\Gamma_{2} \approx \Gamma_{2}^{\text{NLO}} + \Gamma_{\text{cross}}$. Here, $\Gamma_{2}^{\text{NLO}}$ corresponds to a $1/N$ expansion of $\Gamma_{2}$ in the number of real field components in the original action (N=2 for the considered model) to next-to-leading order (NLO)
\begin{equation}
\begin{split}
\Gamma_2^{\text{NLO}} = 2 \cdot \begin{gathered}
\includegraphics[width=2.5cm]{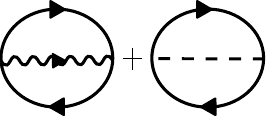}
\end{gathered}
\end{split} \ .
\end{equation}
These diagrams correspond to bosons scattering with mobile dipoles and a dynamical screening of local interactions as illustrated in Fig.~\ref{fig:coverIllustration} (a). Large-N expansions provide a systematic, non-perturbative approximation which does not rely on the weakness of any interaction strength as it contains diagrams of arbitrary order in $J$ and $U$ \cite{berges_introduction_2004,aarts_far--equilibrium_2002, schluenzen_ultrafast_2020}. They have been used to study non-equilibrium dynamics in a variety of systems such as relativistic $O(N)$ theories~\cite{berges_controlled_2001, aarts_nonequilibrium_2006, weidinger_floquet_2016}, ultracold Fermi~\cite{kronenwett_far--equilibrium_2011} and Bose~\cite{gasenzer_ultracold_2009, gasenzer_nonperturbative_2005} gases as well as spin systems \cite{babadi_far--equilibrium_2015, schuckert_non-equilibrium_2018, schuckert_non-local_2020} and give reliable results already for small values of N~\cite{aarts_effective_2008, berges_controlled_2001}. We include an additional cross diagram, the lowest order diagram not contained in $\Gamma_{2}^{\text{NLO}}$, through $\Gamma_{\text{cross}}$ in order to fully capture dynamics up to terms quadratic in interactions and  correctly reproduce the short-time charge dynamics. Within this approximation and due to the rapid loss of memory of the initial state present in our system, the time complexity of our numerics scales linearly with the simulated time and quadratically in the number of lattice sites.

\begin{figure*}[!thb]
\includegraphics[width=\textwidth]{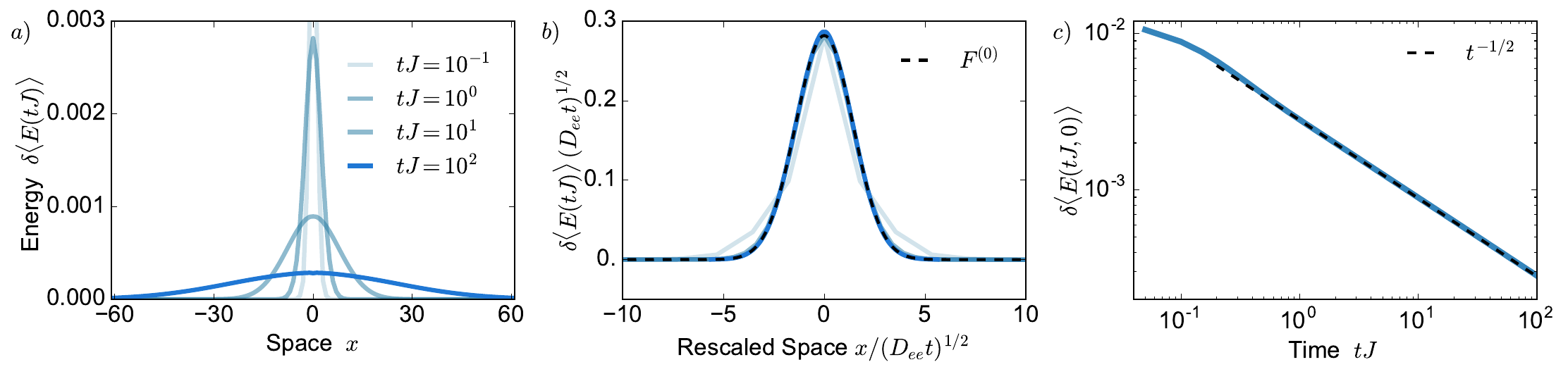}% Here is how to import EPS art
\caption{\label{fig:ScalingE}\textbf{Evolution of non-equilibrium initial states with a homogeneous charge profile and a sharply peaked energy profile.} a) Energy profile at various times. b) The rescaled energy profiles collapse to a Gaussian $F^{(0)}$. Here, the excitation has been normalized to unity. c) The excitation decays algebraically with exponent close to 1/2.}
\end{figure*}

\subsection{Initial States}
Specifying the correlation function $G_{ij}(t_1, t_2)$ at a single point $t_1, t_2 = 0$ corresponds to approximating the initial state as a Gaussian density matrix~\cite{berges_introduction_2004}. In this work, we only consider uncorrelated initial states $G_{ij}(0,0) \propto \delta_{ij}$. Restricting to such a subset significantly reduces numerical effort as the 2-point function $G_{ij}$ remains diagonal in lattice indices for all times due to the form of the interactions.
In order to construct initial states with freely variable charge and energy profiles we also go beyond Gaussian initial states by self-consistently solving the Kadanoff-Baym (KB) equations \eqref{eq:dGamma/DG_mainText}  for the local part of the Hamiltonian $H^{U}$ on each lattice site while enforcing fluctuation-dissipation relations (see App. \ref{app:selfConsistent}). This enables us to represent states $\rho = \bigotimes_{i} \exp \, \{-\beta_{i}(U \hat{n}_{i} (\hat{n}_{i} - 1) - \mu_{i} \hat{n}_{i})\}$ with arbitrary profiles of the inverse temperature $\beta$ and chemical potential $\mu$ fully within the given 2PI approximation. Since states featuring no correlations between different lattice sites carry no energy under the dipole-hopping part of the full Hamiltonian, varying the temperature and chemical potential profiles allows us to independently tune the initial charge and energy profiles. All initial states discussed in this work correspond to  states either at or close to infinite temperature.

\section{Numerical Analysis}
\label{sec:charge_and_energy_dynamics}
\subsection{Coupled Hydrodynamics}
We investigate charge dynamics in the system by preparing Gaussian far-from-equilibrium initial states featuring a homogeneous density profile. On top of this background we create a local particle excitation, leading to a charge density profile of the form $\langle n_{i}\rangle = n_{B} + \delta_{i 0}$. The time evolution of this initial state is shown in Figure \ref{fig:ScalingN}. At times up to $tJ = \mathcal{O}(1)$ we find a quantum coherent regime where the dynamics is dominated by the buildup of correlations between neighboring lattice sites, establishing local equilibrium; \figc{fig:ScalingN}{c}. The numerical results agree with an exact solution of the dynamics to order $\mathcal{O}(t^2)$ (see  App. \ref{app:short_time}).
After this regime, hydrodynamic behaviour emerges with an algebraic decay of the particle excitation. The dynamical exponent $z=4$ indicates subdiffusive transport. We further verify in \figc{fig:ScalingN}{a} that the full spatial profile of the charge density predicted in \eq{eq:1.20} emerges at late times. In particular, the charge density exhibits clear signatures of dipole-conserving hydrodynamics including a non-Gaussian profile featuring dips next to the main peak. More quantitatively, as demonstrated in \figc{fig:ScalingN}{b}, the charge density profile at different times collapses to a scaling form of generalized hypergeometric functions $F^{(1)}$ \footnote{\label{note:F1}\begin{equation*}
F^{(1)}(x) = \frac{1}{\pi}\Gamma\Big(\frac{5}{4}\Big) \,  _{0}F_{2}\Big[\frac{1}{2}, \frac{3}{4}; \frac{x}{256}\Big] - \frac{\sqrt{x}}{8\pi} \Gamma\Big(\frac{3}{4}\Big)  \,  _{0}F_{2}\Big[\frac{5}{4}, \frac{3}{2}; \frac{x}{256}\Big] 
\end{equation*}
where $\Gamma$ denotes the gamma function and $_{p}F_{q}$ the generalized hypergeometric function.}, which has been previously established as a signature of dipole conservation \cite{feldmeier_anomalous_2020, gromov_fracton_2020}.
We point out that the emergence of the subdiffusive dynamical exponent $z=4$ is very robust within our simulations. In fact, we numerically extract exponents in the range of $0.248 - 0.253$ for a variety of initial states with fillings in the range of $n_{\text{B}} = 1  \text{ to } 3$ and interaction strengths ranging from $U = 0 \text{ to } 10$. Both the filling and the interaction strength affect the effective diffusion constants as we discuss below.

\begin{figure*}[t!]
\includegraphics[width=\textwidth]{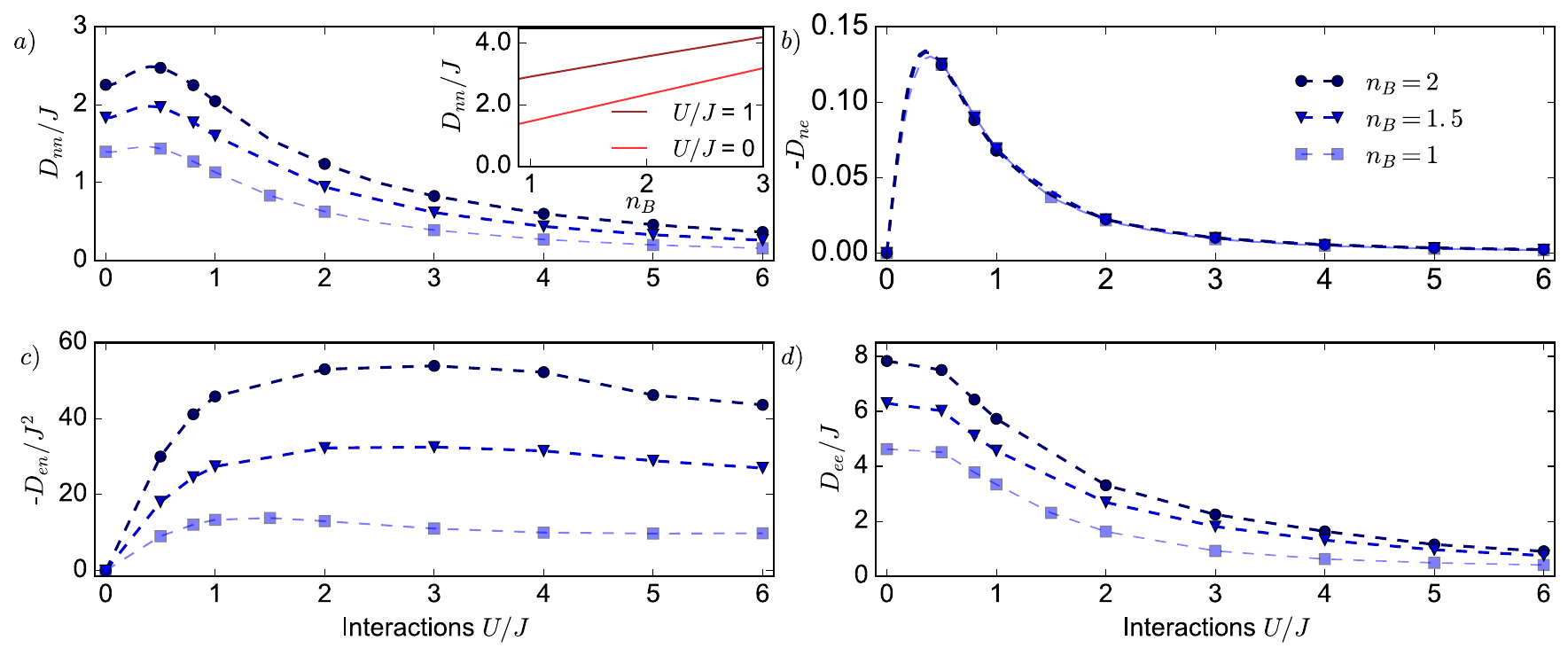}
	\caption{\label{fig:diffusion_Constants}\textbf{Diffusion matrix.} Numerical values of the infinite temperature diffusion coefficients as a function of the local interaction strength $U/J$ and filling fraction $n_{B}$. a) Charge-charge diffusion constant $D_{nn}$ governing the subdiffusive decay of charge excitations.  Inset:  $D_{nn}$ as a function of filling  for two values of $U/J$. b) Charge-energy diffusion constant as computed from Onsager relations. c) Off-diagonal energy-charge diffusion constant. d) Energy-energy diffusion constant. Dashed lines are obtained by applying Eq. \eqref{eq:Onsager_D_ne} to linearly interpolated simulation data.}
\end{figure*}

In the presence of a finite on-site boson-boson repulsion $U > 0$, a charge  excitation generally also leads to an excitation in energy density. In our hydrodynamic description, the dynamics of the charge density are only modified by a renormalization of the diffusion coefficient; see \eq{eq:1.20}. By contrast, the nonzero interaction $U$ qualitatively modifies the dynamics of energy as depicted in Fig.~\ref{fig:coverIllustration} (b). As a nonzero $U$ couples energy to charge density, the subdiffusive charge dominates the dynamics of energy at late times. This confirms the presence of a mixed charge-energy subdiffusive mode as described in \eq{eq:1.20}. 

In order to probe the emergence of a second, diffusive pure energy mode, we prepare states with \textit{homogeneous} filling fraction $n_{B}$ and a highly localized energy inhomogeneity. The time evolution of the energy profile is depicted in Fig. \ref{fig:ScalingE}. The pure-energy excitation decays as $t^{-1/2}$ and exhibits a scaling collapse to a Gaussian $F^{(0)}(x) = \frac{1}{\sqrt{2\pi}} e^{-x^2 / 2}$. These results qualitatively confirm the validity of the coupled hydrodynamic description \eqref{eq:1.20}. Physically, the diffusive pure energy mode, which becomes the only energy mode for $U=0$, is consistent with the diffusive motion of dipoles in the system: While there is no microscopic local dipole density in the system that directly reflects this process, diffusive dipole motion formally manifests itself in a finite dipole-conductivity $\sigma_{dd} = \lim_{\omega \rightarrow 0} \frac{\beta}{2}\braket{j^d(k=0,\omega)j^d(k=0,-\omega)} = C_{nn} \tilde{D}_{nn}/T$ (see App.~\ref{app:cond}).

\subsection{Diffusion Matrix}
\label{sec:diffusion_matrix}

The qualitative hydrodynamic behaviour of the dipole conserving system~\eqref{eq:Hamiltonian} is given in terms of the effective model \eq{eq:1.18}. A full hydrodynamic description further requires knowledge of all diffusion constants. Since diffusion constants are non-universal properties of a quantum system, extracting them requires a direct connection between microscopic and macroscopic physical descriptions, which we obtain through our 2PI effective action approximation. 
We extract the (sub)diffusion constants $\tilde{D}_{nn}$ and $D_{ee}$ according to \eq{eq:1.20} by fitting algebraic curves $c \, t^{-1/z}$ to the late-time behaviour of $\delta\braket{n}(t, 0)$ and $\delta\braket{E}(t, 0)$, the local charge-energy and energy-only excitations. As previously discussed, this robustly yields dynamical exponents of $z = 2$ and $4$ for energy and charge-energy modes, respectively. 

The off-diagonal diffusion constant $D_{en}$ may be obtained by evolving the system to very late times towards its steady state, which is characterized by vanishing currents $j^{d}= j^{e}=0$. Using this condition in the hydrodynamic expansion \eqref{eq:1.16} yields linearly tilted energy and charge profiles in the long time limit, i.e.  $n_{\text{eq}}(x) = n_{B} + t_{n} x$ and $\epsilon_{\text{eq}}(x) = \epsilon_{B} + t_{\epsilon} x$. The ratio of both tilts is related to elements of the diffusion matrix as 
\begin{equation}
\label{eq:tilt_ratios}
    \frac{D_{en}}{D_{\epsilon\epsilon}} = - \frac{t_{\epsilon}}{t_{n}} \ .
\end{equation}
For same-sign tilts of the charge and energy densities, which we observe, this implies $D_{en} < 0$. 
Numerically, we find a parameter dependence of steady state tilt ratios as $D_{en}/D_{ee}=-t_{\epsilon}/t_{n} \approx -4 U n_{B}$. We emphasize that this coincides precisely with the Onsager symmetry \eq{eq:1.22} upon inserting the equilibrium correlations at infinite temperature (see App.~\ref{app:Infinite_T_corr}). We have thus directly shown the emergence of Onsager's relations in our microscopic approach. Accordingly, within our hydrodynamic model, we can use \eq{eq:1.23} to determine the diffusion constants $D_{nn}, D_{ne}$ in terms of the numerically determined $\tilde{D}_{nn}, D_{ee}, D_{en}$. We show all diffusion constants as functions of interaction strength $U$ and for different background fillings in Fig. \ref{fig:diffusion_Constants}.

At constant filling, $D_{ee}$ decays with increasing on-site interactions, in agreement with the intuition that the dynamics should become slower toward the Mott limit of $U \to \infty$, where all many-body eigenstates are completely localized. We note that while the bare charge-charge diffusion constant $D_{nn}$ shows a slight increase at small values of $U/J$ in \figc{fig:diffusion_Constants}{a}, the renormalized diffusion constant $\tilde{D}_{nn}$ decreases monotonically with increasing interaction. Furthermore, both off-diagonal diffusion constants vanish at $U = 0$ leading to a decoupling of charge and energy modes in the absence of boson-boson repulsion as expected.

As the filling increases, excitations becomes more mobile and we find both diagonal diffusion constants increase linearly with $n_B$ for $n_B \gtrsim 1$ as seen in the inset of \figc{fig:diffusion_Constants}{a}. This can be understood as increased particle mobility at higher fillings due to Bose-enhancement. 
By contrast, at decreasing filling direct hopping processes can become infeasible, as no particles may be nearby to coordinate movement with. At sufficiently low filling the dynamics must eventually become frozen due to a strong Hilbert space fragmentation into disjoint sectors~\cite{sala_ergodicity-breaking_2020, morningstar_kinetically_2020}. This fragmentation transition cannot be resolved by the methods developed in this work, as exact identities in the product state basis must be reproduced, which is notoriously difficult for field theoretic approaches.

\section{Discussion and outlook}
\label{sec:conclusion}

We have studied the emergent, coupled hydrodynamics of charge and energy in a system of dipole-conserving lattice bosons at infinite temperature. We have confirmed subdiffusive dynamics of the charge density while in the absence of charge inhomogeneities energy spreads diffusively at late times. Our quantum field theoretic results are well captured by an effective hydrodynamic model whose generalized diffusion matrix can be extracted from our numerics as a function of filling-fraction and interaction strength.

We emphasize that our hydrodynamic description is qualitatively independent of microscopic details.
In particular, as noted in previous works on dipole-conserving hydrodynamics~\cite{gromov_fracton_2020,feldmeier_anomalous_2020}, the subdiffusion of charge with $z=4$ is in agreement with an experimental study in a two-dimensional fermionic system in the presence of a tilted potential ~\cite{guardado-sanchez_subdiffusion_2020}. The associated Hamiltonian is given by $\hat{H}=\hat{H}_{\mathrm{FH}} + F \sum_{\boldsymbol{r}} r_x \hat{n}_{\boldsymbol{r}}$, where $\hat{H}_{\mathrm{FH}}$ is the usual Fermi-Hubbard model.
More generally and independent of whether we consider bosons or fermions and 1D or 2D, the center of mass (or dipole moment) in such a tilted setup is expected to be a conserved quantity up to times $\tau \sim \exp(F/t)$, i.e., exponentially long in the tilt strength, by the arguments of prethermalization~\cite{zhang_universal_2020, khemani_localization_2020, Scherg2021aa,kohlert_experimental_2021} ($t$ is the usual single-particle hopping). Within this timescale the dynamics is governed by a dipole-conserving effective Hamiltonian, such as \eq{eq:Hamiltonian}, with correlated hopping strength $J\sim U (t/F)^2$ in a basis obtained from the Schrieffer-Wolff transformation. Both the number of particles as well as their non-tilt energy are conserved densities in this basis. If the prethermal timescale $\tau$ is longer than the local thermalization time of the resulting effective dipole Hamiltonian, the coupled hydrodynamic theory of \eq{eq:1.18} will be applicable to these two modes. Our results should then be viewed as the system's `prethermal hydrodynamics.' Whether the above condition is satisfied might be verified in quantum gas microscopes by measuring the fluctuations of the dipole moment. In addition, it is an interesting open question how the dynamics of the off-diagonal correlated hopping, and thus the energy, could be measured in cold atom quantum simulation experiments.

After the timescale of prethermalization the dipole moment is no longer strictly conserved; tilt energy and non-tilt energy of the effective Hamiltonian will then be converted into one another. Therefore, at the longest times the diffusive non-tilt energy ceases to be a well defined hydrodynamic mode as shown in Ref.~\cite{gromov_fracton_2020}. Nevertheless, in this late-time regime the coarse-grained charge dynamics is still governed by an emergent hydrodynamic description equivalent to the hydrodynamics of dipole-moment conserving systems, leading to a subdiffusive mode with $z=4$~\cite{guardado-sanchez_subdiffusion_2020, gromov_fracton_2020}.
It is an interesting question for future work to determine whether this dynamical crossover has an impact on the value of the subdiffusion constant $\tilde{D}_{nn}$. For future research, it would further be interesting to consider within our non-equilibrium field theory a continuum system in which momentum is conserved in addition to the conservation laws considered here~\cite{glorioso_breakdown_2021,osborne2021infinite,grosvenor2021hydrodynamics}. Moreover, the effect of dissipation could be incorporated in an effective hydrodynamic description, which constitutes another promising research direction building on our studies.
\newline
\newline 
Simulation codes are available on Zenodo \cite{burchards_ansgar_2022_6528597}.

\vspace{\baselineskip}
\textbf{Acknowledgements}. 
We thank A. Lucas and H. Spohn for interesting discussions. 
We acknowledge support from the Deutsche Forschungsgemeinschaft (DFG, German Research Foundation) under Germany’s Excellence Strategy--EXC--2111--390814868, TRR80 and DFG grants No. KN1254/1-2 and No. KN1254/2-1, BMBF EQUAHUMO, the European Research Council (ERC) under the European Union’s Horizon 2020 research and innovation programme (grant agreement No. 851161), as well as the Munich Quantum Valley, which is supported by the Bavarian state government with funds from the Hightech Agenda Bayern Plus.

\newpage
\appendix
\begin{widetext}
\addcontentsline{toc}{section}{Appendix} % Add the appendix text to the document TOC
\addtocontents{toc}{\protect\addtolength{\protect\cftchapnumwidth}{10pt}}
\addtocontents{lof}{\protect\addtolength{\protect\cftfignumwidth}{40pt}}
\addtocontents{lot}{\protect\addtolength{\protect\cfttabnumwidth}{40pt}}
\part{Appendix} % Start the appendix part
\parttoc % Insert the appendix TOC
%\section{Supplementary Material}
\section{Fluctuation-Dissipation Relations}
\label{app:fdrs}
In this section, we derive the Fluctuation-Dissipation Relations (FDRs) \eq{eq:1.21} from the hydrodynamic equation~\eqref{eq:1.18}.
We will show that in combination with the Onsager symmetry relations (see App.~\ref{app:cond}) these FDRs lead to the relations \eqref{eq:1.22} and \eqref{eq:1.23} between different diffusion constants.

We start with the general real-space solutions to the fluctuating hydrodynamic equation \eqref{eq:1.18}, which can be written as
\begin{equation}
\begin{split}
    u_{a}(t, x) = \int_{-\infty}^{\infty} dx' \, G_{ab}(t, x - x') u_{b}(0, x') + \int_{0}^{t} dt' \int_{-\infty}^{\infty} dx' \,  G_{ab}(t - t', x - x') \partial_{x}^{\alpha_b} \xi_{b}(t', x').
\end{split}
\end{equation}
Here, $u(x) = \big( n(x), e(x) \big)^{T}$ is the vector of conserved densities, $\alpha_{b=n,e} = 2, 1$ denotes the number of derivatives in the associated continuity equations and a sum over repeated subscript indices is implied. $G$ corresponds to the hydrodynamic retarded Green's function, which in Fourier space reads
\begin{equation}
    G(t, q) = e^{- k^2 \underline{D}(k) t} \theta(t),
\end{equation}
with the `scale-dependent' diffusion matrix $\underline{D}(k)$ defined in \eq{eq:1.19}.
We introduce the $2\times 2$ matrix $\underline{C}$ of static equilibrium correlations defined as $C_{ab}(x_1 - x_2) = \lim_{t \rightarrow \infty}  \langle u_{a}(t, x_{1}) u_{b}(t, x_{2}) \rangle$. Since contributions from initial configurations vanish at $t \rightarrow \infty$ and the sources satisfy $\langle \xi_{a}(t_1, x_1) \xi_{b}(t_2, x_2)\rangle = B_{a}^2 \delta_{ab} \delta(t_1 - t_2) \delta(x_1 - x_2) $, these correlations  may be written as 
\begin{equation} \label{eq:Appa1}
\begin{split}
      C_{ab}(x_1 - x_2)  = \int_{0}^{\infty} dt' \int_{-\infty}^{\infty} dx' \, \partial_{x_1}^{\alpha_{c}} G_{ac}(t', x_1 - x') \, \partial_{x_2}^{\alpha_{c}} G_{bc}(t', x_2 - x') \, B_{c}^2.
\end{split}
\end{equation}
Their Fourier is then given by
\begin{equation}
\label{eq:Correlations_q}
\begin{split}
    C_{ab}(k) = \int_{0}^{\infty} dt \, \big(e^{-k^2\underline{D}(k) t}\big)_{ac} \big(e^{-k^2\underline{D}(k)t}\big)_{bc} \, B_c^2 \, k^{2\alpha_c} .
\end{split}
\end{equation}
Taking a time derivative immediately yields the FDRs $\underline{D}(k) C(k) + C(k) \underline{D}(k)^{T} = \underline{B}(k)\underline{B}(-k)^T$ of \eq{eq:1.21}.

We notice that, a priori, the static correlations in \eq{eq:Correlations_q} appear to depend on the wavevector $k$, which is equivalent to real space correlations different from delta-function form. However, Onsager symmetry of the kinetic coefficients (see App.~\ref{app:cond}) implies that the matrix $\underline{D}(k)\underline{C}(k)$ is symmetric~\cite{Onsager_1931, spohn_hydro_1996}. Applying this symmetry condition to the explicit form of the correlations that can be obtained from \eq{eq:Correlations_q} leads to static correlations \textit{independent} of wavevector $k$ as expected. Combining the Onsager relations with the above FDRs then leads to the relations \eqref{eq:1.22} and \eqref{eq:1.23} of the main text. In particular, we can obtain the coefficient $D_{ne}$ entering the diffusion matrix $\underline{D}(k)$ through the relation
\begin{equation}
\label{eq:Onsager_D_ne}
    D_{ne} = - \frac{C_{en} \tilde{D}_{nn}}{C_{ee} + C_{en} D_{en} / D_{ee}},
\end{equation}
expressed entirely in terms of constants or correlations that can be extracted directly from our numerical method (at infinite temperature).

\section{Linear Response Conductivities From Non-Equilibrium transport\label{app:cond}}

In the main text, we have calculated transport coefficients from non-equilibrium quenches. Here we bring these results into contact with transport considered within linear response. Of central importance is the conductivity matrix, whose components in the high-temperature limit can be written as~\cite{bertini_finite-temperature_2021}
\begin{equation}
\sigma_{ab}(k,\omega)=\frac{\beta}{2} \braket{j^a(k,\omega)j^b(-k,-\omega)}_\beta.
\end{equation}
Here, $j^{a=n,e}$ are the current densities of charge (which is obtained from the dipole current through a derivative according to \eq{eq:1.14}) and energy and $\beta$ is the inverse temperature. 
In the following, we derive these conductivities from the solution of the coupled hydrodynamics in Eq.~\eqref{eq:1.20}. Onsager symmetry then amounts to requiring a symmetric conductivity matrix and will be shown to be equivalent to \eq{eq:1.22}.

To begin, we note that due to the continuity equations, 
\begin{equation}
\braket{j^a(k,\omega), j^b(-k,-\omega)}_\beta=\frac{\omega^2}{k^2}\braket{u_a(k,\omega) u_b(-k,-\omega)}_\beta = \frac{\omega^2}{k^2} C_{ab}(k,\omega),
\end{equation}
where the $u_{a=n,e}$, are again the densities corresponding to the conserved densities.

We consider then a density profile $\delta \braket{u_a(k,t=0)}$ created via an adiabatic switch-on of a perturbation at times $t<0$. For $t>0$, the external perturbation is switched off and the system relaxes back to equilibrium. It can be shown from linear response (see e.g. book by Forster \cite{Forster}) that the Laplace transform $\delta\braket{u_a(k,z)}=\int_0^\infty \, dt e^{izt} \delta\braket{u_a(k,t)}$ of the time evolved density perturbations $\delta\braket{u_a(k,t)}=\braket{u_a(k,t)}-\braket{u_a(k)}_{\mathrm{eq}}$ is given by
\begin{equation}
\delta\braket{u_a(k,z)} = \beta [S(k,z)\chi^{-1}(k)]_{ab} \delta \braket{u_a(k,t=0)}, \label{eq:LR_pert}
\end{equation}
where $\chi(k)$ is the matrix of static susceptibilities at inverse temperature $\beta$ and $S$ is the Kubo correlation function. $\chi(k)$ and $S$ are connected to the imaginary part of the susceptibility by $\chi(k)=\int d\omega \chi''(k,\omega)/(\pi\omega)$ and $\mathrm{Im}(S(k,\omega+i\epsilon))=\chi^{''}(k,\omega)/(\beta\omega)$. 
Linear response written as in Eq.~\ref{eq:LR_pert} can be directly compared to our long-wavelength solution obtained from linear fluctuating hydrodynamics in Eq.~\ref{eq:1.20}. The Laplace transform of this solution is given by
\begin{equation}
\begin{pmatrix}
n_k(z) \\ e_k(z)
\end{pmatrix}= \begin{pmatrix}
\frac{i}{z+i\tilde D_{nn}k^4} & 0 \\
\frac{D_{en}}{D_{ee}}\left(\frac{i}{z+iD_{ee}k^2}-\frac{i}{z+i\tilde D_{nn}k^4}\right) & \frac{i}{z+iD_{ee}k^2}
\end{pmatrix}\begin{pmatrix}
n_k(t=0) \\ e_k(t=0)
\end{pmatrix}.
\end{equation}
Hence, the matrix $M$ on the right hand side corresponds to $\beta S(k,z)\chi^{-1}(k)$. Furthermore, we use the classical fluctuation-dissipation relation $C(k,\omega)= 2\chi^{''}(k,\omega)/(\beta\omega)$, which implies $\chi(k)=\beta C(k)$ for the static susceptibility. Assembling all of the above relations, we finally obtain the conductivity matrix as
$\sigma(k,\omega)=\frac{\omega^2}{k^2} \mathrm{Im}(M(k,w+i\epsilon)) C(k)$, or more explicitly
\begin{equation}
\sigma(k,\omega)=\beta\begin{pmatrix}
\frac{C_{nn}\tilde D_{nn}(\omega k)^2}{\omega^2+(\tilde D_{nn}k^4)^2} & \frac{C_{ne}\tilde D_{nn}(\omega k)^2}{\omega^2+(\tilde D_{nn}k^4)^2}\\
\frac{D_{en}C_{nn}}{D_{ee}}\left(\frac{D_{ee}\omega ^2}{\omega^2+(D_{ee}k^2)^2}-\frac{\tilde D_{nn}(\omega k)^2}{\omega^2+(\tilde D_{nn}k^4)^2} \right)+\frac{C_{en}D_{ee}\omega^2}{\omega^2+(D_{ee}k^2)^2} & \frac{D_{en}C_{ne}}{D_{ee}}\left(\frac{D_{ee}\omega ^2}{\omega^2+(D_{ee}k^2)^2}-\frac{\tilde D_{nn}(\omega k)^2}{\omega^2+(\tilde D_{nn}k^4)^2} \right)+\frac{C_{ee}D_{ee}\omega^2}{\omega^2+(D_{ee}k^2)^2}
\end{pmatrix},
\end{equation}
where we suppressed the $k$ dependence of the static correlations (at $\beta=0$, $C(k)=C$ exactly as shown below).
In particular, we find for the zero momentum conductivity in the static limit
\begin{equation} \label{eq:AppB1}
\lim_{\omega\rightarrow 0}\sigma(k=0,\omega)=\beta\begin{pmatrix}
0 & 0 \\
C_{nn}D_{en}+D_{ee}C_{en} & C_{ne}D_{en}+D_{ee}C_{ee}.
\end{pmatrix},
\end{equation}
which gives the generalized Einstein relations for the diffusive energy mode and a vanishing conductivity for the subdiffusive charge density. From \eq{eq:AppB1} we thus find that demanding the conductivity matrix to be symmetric leads to \eq{eq:1.22} of the main text, i.e.
\begin{equation}
C_{nn}D_{en}+C_{en}D_{ee} = 0.
\end{equation}
Inserting this condition back into the FDRs of \eq{eq:1.20} then yields \eq{eq:1.23}, i.e. $C_{en}D_{nn}+C_{ee}D_{ne}=0$, which further implies that $\underline{D}(k)\underline{C}$ is a symmetric matrix.

Finally, we notice that we can formally also define a `dipole conductivity' via the correlations of the dipole current $j^d$, $\sigma_{dd} := \lim_{\omega\rightarrow 0}\lim_{k\rightarrow 0}\frac{\beta}{2}\braket{j^d(k,\omega)j^d(-k,-\omega)}$. 
Evaluating this expression yields a finite dipole conductivity $\sigma_{dd} = \beta C_{nn} \tilde D_{nn}$, consistent with a diffusive motion of dipoles through the system.

\section{Infinite-Temperature Equilibrium Correlations}
\label{app:Infinite_T_corr}
The static infinite temperature correlators $C_{ab}$ can be calculated analytically. The static connected correlations are defined as 
\begin{equation}
    C_{nn} = \langle \hat{n}_{i} \hat{n}_{i} \rangle ,  \quad 
    C_{ne} = C_{en} =  \langle \hat{e}_{i} \hat{n}_{i} \rangle , \quad
    C_{ee} = \langle \hat{e}_{i} \hat{e}_{i} \rangle 
\end{equation}
with energy-density operator $\hat{e}_{i} = h^{U}_{i} + \Big( \frac{1}{4}(\hat{h}^{D}_{i-1} + \hat{h}^{D}_{i-1}) + \frac{1}{2}\hat{h}^{D}_{i} + h.c.\Big)$.

Exploiting the fact that the infinite temperature density matrix is given as $\rho = 1/\mathcal{N} e^{-\kappa n}$ with $\kappa = -\beta \mu$ fixed by the particle expectation value $\langle n \rangle = n_{B}(\kappa) = (e^{\kappa} - 1)^{-1}$ and $\mathcal{N}$ a normalization constant we obtain the static correlations
\begin{equation}
    C_{nn} = n_{B}^2 + n_{B},
    \qquad 
    C_{en} = U (4 n_{B}^3 + 4 n_{B}^2),
    \qquad
    C_{ee} = 4 U^{2} (5 n_{B}^4 + 6 n_{B}^3 + n_{B}^2 ) + 4 J^{2} n_{B}^2 (n_{B}^2 + 2 n_{B} + 1) \ .
\end{equation}

\section{Equations of Motion from 2PI Effective Action}
\label{app:eoms}
\subsubsection{Constructing the Non-Equilibrium Partition Function}
The action on the Schwinger-Keldysh contour corresponding to the Hamiltonian (\ref{eq:Hamiltonian}) with normalization $J = 1$ is 

\begin{equation}
S[\phi] = \int_{\mathcal{C}}  \overline{\phi}_j i \partial_{t_1} \phi_j - \overline{\phi}_{i} \phi_{j} V_{ijkl} \overline{\phi}_{k}\phi_{l} - U \overline{\phi}_{i}^2 \phi_{i}^2
\end{equation}
with a symmetrized interaction tensor

\begin{equation}
V_{ljkm} = \frac{1}{2} \big ( \delta_{l k j+1 m-1} + \delta_{l k j-1 m+1}  
+\delta_{j m k+1 l-1}  + \delta_{j m k-1 l+1} \big )
\end{equation}
where $\delta_{ijkl}$ is equal to unity if and only if all index values are identical. 
\begin{figure}[!htb]
	\includegraphics[width=.5\textwidth]{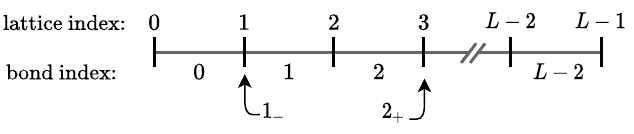}
	\caption{Labeling of lattice sites and bonds in a system of size L. Lattice sites and bonds are numbered from left to right beginning with 0. The first lattice site may also be labelled as $1_{-}$ since it is to the left of the first bond, similarly $2_{+}$ refers to the third lattice site.}
	\label{fig:BondIndices}
\end{figure}
Due to the fact that the tensor $V$ vanishes whenever the first or last two indices denote non-consecutive lattice sites it is useful to introduce an additional bond-index denoting ordered, consecutive pairs of lattice sites. Formally we define these bond-indices as tuples of consecutive lattice sites $(i, i\pm1)$ and consider them left/ rightward oriented if the second lattice index is smaller/larger than the right one. For clarity we denote the bond indices by lower-case Greek letters in summations, marking with a dot rightward oriented bond-indices while lattice-indices will be denoted by lower-case Latin letters. Further, the lattice sites to the right and left of a bond are denoted by the bond index subscripted with a $+$ and $-$ sign respectively. For example if $\sigma = 4$ denotes the fourth bond of the system from the left then $\sigma_{-}$ denotes the fourth lattice site while $\sigma_{+}$ denotes the fifth. This is illustrated in Fig. \ref{fig:BondIndices}.Using this notation the tensor V can be expressed in the form

\begin{equation}
\begin{split}
V_{\alpha \dot{\mu}}  & =  V_{\dot{\alpha} \mu} = \frac{1}{2}(\delta_{\alpha \mu + 1} + \delta_{\alpha \mu - 1})\\
V_{\alpha \mu} & =  V_{\dot{\alpha} \dot{\mu}} = 0 
\end{split}
\end{equation}
and the partition function reads

\begin{equation}
\mathcal{Z} =  \int D\phi \, \text{exp} \, i \ \Big(  \int_{\mathcal{C}} \overline{\phi}_j i \partial_t \phi_j - 2 \overline{\phi}_{\sigma_{-}} \phi_{\sigma_{+}} V_{\sigma \dot{\mu}} \overline{\phi}_{\mu_{-}} \phi_{\mu_{+}} - U \overline{\phi}_{i}^2 \phi_{i}^2 + \text{sources}\Big) \ .
\end{equation}
We continue by decoupling both quartic terms in the action via introduction of a complex-valued  Hubbard-Stratonovich field $\chi$ localized on the bonds as well as a real-valued Hubbard-Stratonovich field $\Delta$ localized on the lattice, obtaining the final action 

\begin{equation}
\begin{split}
S[\phi, \chi, \Delta] = & \int_{\mathcal{C}}  \overline{\phi}_i i \partial_t \phi_i + \overline{\chi}_{\dot{\sigma}} V^{-1}_{\dot{\sigma} \mu}\chi_{\mu} +  \frac{1}{2}\Delta_{i} U^{-1}_{i} \Delta_{i} \\ & - \sqrt{2} \big (\overline{\chi}_{\dot{\alpha}} \overline{\phi}_{\alpha_{-}} \phi_{\alpha_{+}}  +  \overline{\phi}_{\alpha_{+}} \phi_{\alpha_{-}} \chi_{\alpha } + \overline{\phi}_{i} \phi_{i} \Delta_{i} \big )
\end{split} 
\end{equation}

\begin{equation}
\text{sources} \rightarrow \text{sources} + \overline{\chi}_{\dot{\sigma}} J^{\chi}_{\dot{\sigma}}
+  \overline{J}^{\, \chi}_{\sigma}\chi_{\sigma} + \overline{\chi}_{\dot{\sigma}} R^{\chi}_{\dot{\sigma} \mu}\chi_{\mu} + \Delta_{i} J^{\Delta}_{i} + \frac{1}{2} \Delta_{i} R^{\Delta}_{ij}\Delta_j \ .
\end{equation}
Formally this transformation is achieved by multiplication of the partition function with the two unities 

\noindent\begin{minipage}{.5\linewidth}
\begin{equation*}
1 = \int \, d\,\overline{\chi}^{\, 0}d\chi^{0} \, \text{exp } i \big(\int_{\mathcal{C}} \overline{\chi}^{\, 0}_{\dot{\sigma}} V^{-1}_{\dot{\sigma} \mu}\chi^{0}_{\mu} \big )
\end{equation*}
\end{minipage}%
\begin{minipage}{.5\linewidth}
\begin{equation}
1 = \int d \Delta^{0} \, \text{exp } i \big (\int_{\mathcal{C}} \frac{1}{2}\Delta^{0}_{i} U^{-1}_{i} \Delta^{0}_{i} \big )
\end{equation}
\end{minipage}
and consecutive field-shifts

\noindent\begin{minipage}{.5\linewidth}
\begin{equation}
\overline{\chi}^{\, 0}_{\dot{\sigma}} \rightarrow \overline{\chi}_{\dot{\sigma}} - \sqrt{2} \, \overline{\phi}_{\alpha_{+}} \phi_{\alpha_{-}} V_{\alpha \dot{\sigma}} 
\end{equation}
\begin{equation}
\chi^{0}_{\sigma} \rightarrow \chi_{\sigma} - \sqrt{2} \, \overline{\phi}_{\alpha_{-}} \phi_{\alpha_{+}} V_{\dot{\alpha}\sigma} 
\end{equation}
\end{minipage}%
\begin{minipage}{.5\linewidth}
\begin{equation}
\Delta^{0}_{i} \rightarrow \Delta_{i} - \sqrt{2}\, \overline{\phi}_{i} \phi_{i} U_{i} \ .
\end{equation}
\end{minipage}
\newline

\subsubsection{Causal Equations of Motion via 2PI Effective Action}
The 2PI effective action is obtained as the Legendre transform of the connected generating functional $W = i \log \mathcal{Z}$ with respect to the local source terms $J, J^{\Delta}, J^{\chi}$ as well as the non-local sources $R, R^{\Delta}, R^{\chi}$. It may be regarded as a functional of the field expectation values $\varphi_{i}(t) = \langle \phi_{i}(t) \rangle, \, X_{\sigma}(t) = \langle \chi_{\sigma}(t) \rangle, \, N_{i}(t) = \langle \Delta_{i}(t) \rangle$ and the full, connected, non-equilibrium field correlators $G_{ij}(t_1, t_2) = \langle \overline{\phi}_{i}(t_1) \phi_{j}(t_2) \rangle$ , \, $D_{\dot{\sigma}\mu}(t_1, t_2) =  \langle \overline{\chi}_{\dot{\sigma}}(t_1) \chi_{\mu}(t_2) \rangle$, \, $O_{ij}(t_1, t_2) =  \langle \Delta_{i}(t_1) \Delta_{j}(t_2) \rangle$ on the Schwinger-Keldysh contour.
The 2PI effective action may be split into up to 1-loop contributions and higher order contributions contained in $\Gamma_2$ as  follows
\begin{align}
\label{eq:Gamma1loop}
    \Gamma^{\text{2PI}}[\phi, \chi, \Delta, G, D, O] = \, & S[\phi, \chi, \delta] + i \text{ tr}\log G^{-1} + i \text{ tr } G_{0}^{-1, T} G +  i \text{ tr} \log D^{-1} + i \text{ tr} D_{0}^{-1, T} D \notag\\ & +   \frac{i}{2} \text{ tr } \log O^{-1} + \frac{i}{2} \text{ tr } O_{0}^{-1} O + \Gamma_2
\end{align}
where the free propagators have been defined as 
\begin{equation}
    i G_{0, ij}(t_1, t_2) = \frac{\delta^{2}S[\phi, \chi, \Delta]}{\delta \overline{\phi}_{j}(t_2)\delta \phi_{i}(t_1)} 
\end{equation}
\begin{equation}
    i D_{0, \dot{\sigma}\mu}(t_1, t_2) = \frac{\delta^{2}S[\phi, \chi, \Delta]}{\delta \overline{\chi}_{\mu}(t_2)\delta \chi_{\sigma}(t_1)} 
\end{equation}
\begin{equation}
    i O_{0, ij}(t_1, t_2) = \frac{1}{2} \frac{\delta^{2}S[\phi, \chi, \Delta]}{\delta \overline{\Delta}_{i}(t_1)\delta \Delta_{j}(t_2)} \ .
\end{equation}
Since the macroscopic fields $\phi$ and  $\chi$ vanish at all times for the initial states considered in this text and all information contained in the field expectation value $N_{i}$ may be reexpressed in terms of $G(t, t)$, we drop them in subsequent equations.
At this point, the equations of motion may be determined explicitly from the stationarity conditions
\begin{equation}
\label{eq:dGamma/dG}
\frac{\delta \Gamma ^{\text{2PI}}[G, D, O]}{\delta G_{ij}(t_1, t_2)}  = 0 \ ,
\qquad \qquad
\frac{\delta \Gamma ^{\text{2PI}}[G, D, O]}{\delta D_{\dot{\sigma}\mu}(t_1,t_2)}  = 0 \ ,
\qquad \qquad
\frac{\delta \Gamma ^{\text{2PI}}[G, D, O]}{\delta O_{ij}(t_1, t_2)}  = 0 \  .
\end{equation}
Defining self-energies for each correlator by 
\begin{equation}
    \Sigma_{ij}(t_1, t_2) = \frac{\delta\Gamma_{2}}{\delta G_{ji}(t_2, t_1)} \ ,
\qquad \qquad
    \Pi_{\dot{\sigma}\mu}(t_1, t_2) = \frac{\delta\Gamma_{2}}{\delta D_{\mu \dot{\sigma}}(t_2, t_1)} \ ,
\qquad \qquad
    \Omega_{ij}(t_1, t_2) = \frac{\delta\Gamma_{2}}{\delta O_{ij}(t_1, t_2)} \ ,
\end{equation}
the equations of motion can be rewritten in integral form

\begin{equation}
\label{eq:dGamma/dG_integral}
    \partial_{t_1}G_{ij}(t_1, t_2) = \delta_{\mathcal{C}}(t_1, t_2) \delta_{ij}- \int_{\mathcal{C}} dt \,  \Sigma_{il}(t_1, t) G_{lj}(t, t_2) 
\end{equation}

\begin{equation}
\label{eq:dGamma/dD_integral}
    D_{\dot{\sigma}\mu}(t_1, t_2) = i V_{\dot{\sigma}\mu}\delta_{\mathcal{C}}(t_1, t_2) + i V_{\dot{\sigma}\alpha} \int_{\mathcal{C}} dt \, \Pi_{\alpha \dot{\beta}}(t_1, t) D_{\dot{\beta} \mu}(t, t_2) 
\end{equation}

\begin{equation}
\label{eq:dGamma/dO_integral}
    O_{ij}(t_1, t_2) = i U_{i} \delta_{ij}\delta_{\mathcal{C}}(t_1, t_2) -  i U_{i} \int_{\mathcal{C}} dt \,  \Omega_{il}(t_1, t) O_{lj}(t, t_2) \ .
\end{equation}
As discussed in the main text, $\Gamma_2$ is equal to the sum over all closed, 2-particle irreducible Feynman diagrams. We employ the approximation $\Gamma_{2} = \Gamma_{2}^{\text{NLO}} + \Gamma_{2}^{\text{cross}}$ with
\begin{equation}
\begin{split}
\Gamma_2^{\text{NLO}} = 2\cdot \begin{gathered}
\includegraphics[width=2.5cm]{Gamma_NLO.pdf}
\end{gathered}
 = 2i & \int_{\mathcal{C}} dt_{1} dt_{2} \, V_{\dot{\sigma} \kappa} D_{\kappa \dot{\nu}}(t_1, t_2) V_{\dot{\nu}\mu} G_{\mu_{-} \sigma_{-}}(t_2, t_1) G_{\sigma_{+}\mu_{+}}(t_1, t_2) \\ & + i \int_{\mathcal{C}} dt_{1} dt_{2} \, U_{i}^{2} G_{ii}(t_1, t_2) G_{ii}(t_2, t_1) O_{ii}(t_1, t_2) 
\end{split}
\end{equation}

\begin{equation}
\begin{split}
\Gamma_2^{\text{cross}} = 
\begin{gathered}
\includegraphics[height=1.2cm]{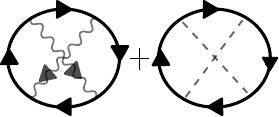}
\end{gathered} = & \, 2 i \iiiint_{\mathcal{C}} dt_{1 \dots 4} \, V_{\dot{\sigma}\mu} V_{\dot{\nu}\kappa} G_{\sigma_{-} \nu_{+}}(t_4, t_3) G_{\nu_{-} \mu_{+}}(t_3, t_2) G_{\mu_{-} \kappa_{+}}(t_2, t_1) G_{\kappa_{-} \sigma_{+}}(t_1, t_4) \\ & + i \iiiint_{\mathcal{C}} dt_{1 \dots 4} \, U_{i}^{2} G_{ii}(t_1, t_2) G_{ii}(t_2, t_1) G_{ii}(t_1, t_2) G_{ii}(t_2, t_1)
\end{split}
\end{equation}
Where we have made use of the Feynman rules
\begin{equation}
- \sqrt{2} i V_{\dot{\mu}  \sigma}  = 
\begin{gathered}
\includegraphics[height=2cm]{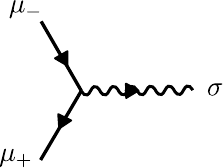}
\end{gathered}
\qquad \qquad
- \sqrt{2} i V_{\mu \dot{\sigma}}  =  
\begin{gathered}
\includegraphics[height=2cm]{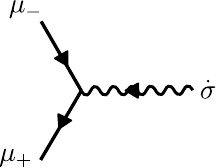}
\end{gathered}
\qquad \qquad
- \sqrt{2} i U_{l} \delta_{l j k}  = 
\begin{gathered}
\includegraphics[height=2cm]{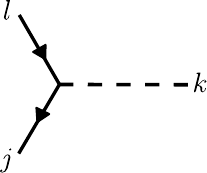}
\end{gathered}
\end{equation}
\begin{equation}
    G_{ij}(t_1, t_2) = \begin{gathered}
\includegraphics[width=2cm]{Propagator_G.pdf}
\end{gathered}
\qquad \qquad
D_{\dot{\sigma} \mu}  = 
\begin{gathered}
\includegraphics[width=2cm]{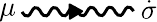}
\end{gathered}
\qquad \qquad
O_{ij}  = 
\begin{gathered}
\includegraphics[width=2cm]{Propagator_O.pdf}
\end{gathered} 
\end{equation}

\begin{equation}
D_{0, \dot{\sigma} \mu}  = 
\begin{gathered}
\includegraphics[width=2cm]{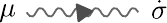}
\end{gathered}
\qquad \qquad
O_{0, ij}  = 
\begin{gathered}
\includegraphics[width=2cm]{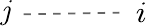}
\end{gathered}\ .
\end{equation}
Additionally, we split off a time-local part from $G, \Sigma, \Pi$ and $ \Omega$  as follows
\begin{equation}
A(t_1, t_2) = -i \delta_{\mathcal{C}}(t_1 - t_2) A^{(0)}(t_1) + i \overline{A}(t_1, t_2)
\end{equation}
with a slightly different decomposition for the propagators $D$ and $O$
\begin{equation}
D(t_1, t_2) = -i \delta_{\mathcal{C}}(t_1 - t_2) D^{(0)}(t_1) + 2 i V \overline{D}(t_1, t_2) V\, 
\end{equation}

\begin{equation}
O(t_1, t_2) = -i \delta_{\mathcal{C}}(t_1 - t_2) O^{(0)}(t_1) + 2 i U \overline{O}(t_1, t_2) U\ .
\end{equation}
The non-local parts of propagators and self-energies are further decomposed into statistical and spectral components
\begin{equation}
 \overline{A}(t_1, t_2)= A^{F}(t_1, t_2) - \frac{i}{2} A^{\rho}(t_1, t_2) \text{sgn}_{\mathcal{C}}(t_1 - t_2) \ .
\end{equation}
We note that the time-local contributions $G^{(0)}, \Pi^{(0)}$ and $\Omega^{(0)}$ vanish and that the statistical and spectral components of $G$ are denoted by $F, \rho$ instead of $G^{F}, G^{\rho}$.
The above decomposition removes explicit reference to the closed time contour since the resulting spectral and statistical components are independent of contour part.

\subsubsection{Full Set of Evolution Equations}
Making use of these decompositions and Eqs. (\ref{eq:dGamma/dG_integral} - \ref{eq:dGamma/dO_integral}) we obtain the final and explicitly causal equations of motion for the 2-point Green's function
\begin{equation}\label{eq:dtF}
\begin{split}
\partial_{t_1} F_i(t_1, t_2) = i \Big  ( & \Sigma^{(0)}_i(t_1) F_i(t_1, t_2)+ \int_{t_0}^{t_1} dt \ \Sigma^{\rho}_i(t_1, t) F_i(t, t_2) \\ -  &  \int_{t_{0}}^{t_2} dt \ \Sigma^{F}_i(t_1, t) \rho_i(t, t_2) \Big)
\end{split}
\end{equation}

\begin{equation}
\label{eq:dtP}
\partial_{t_1} \rho_i(t_1, t_2) = i \Big( \Sigma^{(0)}_i(t_1) \rho_i(t_1, t_2)+ \int_{t_2}^{t_1} dt \ \Sigma^{\rho}_i(t_1, t) \rho_i(t, t_2) \Big) \ ,
\end{equation}
the corresponding self-energies
\begin{equation}
\label{eq:Sigma_loc}
\Sigma^{(0)}_i(t) = 4 U_i \big( F_i(t, t) - \frac{1}{2}\big)
\end{equation}
\begin{equation}\label{eq:Sigma_F}
\begin{split}
\Sigma^{\text{NLO}, F}_l(t_1, t_2)  = \quad \qquad \qquad &  \\  -  4i  \Big\{ U_{l}^{2} F_l(t_1, t_2) O^{F}(t_1, t_2)_l  & + V_{\dot{l-1}\mu} V_{l-1 \dot{\sigma}}D^{F}_{\mu\dot{\sigma}} (t_1, t_2)F_{l-1} (t_1, t_2)   + V_{l\dot{\mu}} V_{\dot{l} \sigma} D^{F}_{\dot{\mu} \sigma} (t_1, t_2) F_{l+1}(t_1, t_2)  \\   -\frac{1}{4} \Big( U_{l}^{2} \rho_l(t_1, t_2) O^{\rho}_l(t_1, t_2)  & + V_{\dot{l-1}\mu} V_{l-1 \dot{\sigma}}D^{\rho}_{\mu\dot{\sigma}} (t_1, t_2)\rho_{l-1} (t_1, t_2)  + V_{l\dot{\mu}} V_{\dot{l} \sigma} D^{\rho}_{\dot{\mu} \sigma} (t_1, t_2) \rho_{l+1}(t_1, t_2)\Big) \Big\}
\end{split}
\end{equation}

\begin{equation}\label{eq:Sigma_P}
\begin{split}
\Sigma^{\text{NLO}, \rho}_l(t_1, t_2) = \quad \qquad \qquad & \\ 4 i  \Big\{ U_{l}^{2}\rho_l(t_1, t_2)  O^{F}_l(t_1, t_2)  & + V_{\dot{l-1}\mu} V_{l-1 \dot{\sigma}} D^{\rho}_{\mu\dot{\sigma}}(t_1, t_2)  F_{l-1} (t_1, t_2) + V_{l\dot{\mu}} V_{\dot{l} \sigma} D^{F}_{\dot{\mu} \sigma}(t_1, t_2)   \rho_{l+1}(t_1, t_2)  \\  + U_{l}^{2} F_l (t_1, t_2) O^{\rho}_l(t_1, t_2)  &  + V_{\dot{l-1}\mu} V_{l-1 \dot{\sigma}} D^{F}_{\mu\dot{\sigma}}(t_1, t_2)  \rho_{l-1} (t_1, t_2) + V_{l\dot{\mu}} V_{\dot{l} \sigma} D^{\rho}_{\dot{\mu} \sigma}(t_1, t_2)  F_{l+1}(t_1, t_2) \Big\} 
\end{split}
\end{equation}

\begin{equation}\label{eq:Sigma_F_cross}
\begin{split}
\Sigma^{\text{cross}, F}_l(t_1, t_2)  = \quad \qquad \qquad &  \\  -  4i  \Big\{ U_{l}^{2} F_l(t_1, t_2) \Omega^{F}(t_1, t_2)_l  & + V_{\dot{l-1}\mu} V_{l-1 \dot{\sigma}} \Pi^{F}_{\mu\dot{\sigma}} (t_1, t_2)F_{l-1} (t_1, t_2)   + V_{l\dot{\mu}} V_{\dot{l} \sigma} \Pi^{F}_{\dot{\mu} \sigma} (t_1, t_2) F_{l+1}(t_1, t_2)  \\   -\frac{1}{4} \Big( U_{l}^{2} \rho_l(t_1, t_2) \Omega^{\rho}_l(t_1, t_2)  & + V_{\dot{l-1}\mu} V_{l-1 \dot{\sigma}}\Pi^{\rho}_{\mu\dot{\sigma}} (t_1, t_2)\rho_{l-1} (t_1, t_2)  + V_{l\dot{\mu}} V_{\dot{l} \sigma} \Pi^{\rho}_{\dot{\mu} \sigma} (t_1, t_2) \rho_{l+1}(t_1, t_2)\Big) \Big\}
\end{split}
\end{equation}

\begin{equation}\label{eq:Sigma_P_cross}
\begin{split}
\Sigma^{\text{cross}, \rho}_l(t_1, t_2) = \quad \qquad \qquad & \\ 4 i  \Big\{ U_{l}^{2}\rho_l(t_1, t_2)  \Omega^{F}_l(t_1, t_2)  & + V_{\dot{l-1}\mu} V_{l-1 \dot{\sigma}} \Pi^{\rho}_{\mu\dot{\sigma}}(t_1, t_2)  F_{l-1} (t_1, t_2) + V_{l\dot{\mu}} V_{\dot{l} \sigma} \Pi^{F}_{\dot{\mu} \sigma}(t_1, t_2)   \rho_{l+1}(t_1, t_2)  \\  + U_{l}^{2} F_l (t_1, t_2) \Omega^{\rho}_l(t_1, t_2)  &  + V_{\dot{l-1}\mu} V_{l-1 \dot{\sigma}} \Pi^{F}_{\mu\dot{\sigma}}(t_1, t_2)  \rho_{l-1} (t_1, t_2) + V_{l\dot{\mu}} V_{\dot{l} \sigma} \Pi^{\rho}_{\dot{\mu} \sigma}(t_1, t_2)  F_{l+1}(t_1, t_2) \Big\} \ ,
\end{split}
\end{equation}
where $\Sigma^{F/\rho} = \Sigma^{\text{NLO}, F/\rho} + \Sigma^{\text{cross}, F/\rho}$.
The Hubbard-Stratonovich field mediating on-site interactions
\begin{equation}
\label{eq:O_loc}
O_{ij}^{(0)} = - U_{i} \delta_{ij}
\end{equation}

\begin{equation}
\label{eq:O_F}
\begin{split}
O^{F}_{ij}(t_1, t_2) = & \ \frac{i}{2} \Omega^{F}_j(t_1, t_2) \delta_{ij}  +  U_{i} \Big(\int_{0}^{t_1}dt \ \Omega^{\rho}_i(t_1, t) O^{F}_{ij}(t, t_2) \\ & -\int_{0}^{t_2}dt \ \Omega^{F}_i(t_1, t) O^{\rho}_{ij}(t, t_2) \Big )
\end{split}
\end{equation}

\begin{equation}
\label{eq:O_P}
O^{ \rho}_{ij}(t_1, t_2) = \frac{i}{2} \Omega^{\rho}_j(t_1, t_2)\delta_{ij} + U_i \int_{t_2}^{t_1}dt \ \Omega^{\rho}_i(t_1, t) O^{\text{NLO}, \rho}_{ij}(t, t_ 2) \ ,
\end{equation}
its corresponding self-energies
\begin{equation}
\label{eq:Om_F}
\Omega^{F}_j(t_1, t_2) = -2 \Big (F_j(t_1, t_2) F_{j}^{*}(t_1, t_2) - \frac{1}{4} \rho_j(t_1, t_2) \rho_{j}^{*}(t_1, t_2)\Big)
\end{equation}

\begin{equation}
\label{eq:Om_P}
\Omega^{\rho}_j(t_1, t_2) = -2 \Big(\rho_{j}(t_1, t_2) F_{j}^{*}(t_1, t_2) + F_j(t_1, t_2) \rho_{j}^{*}(t_1, t_2)\Big) \ ,
\end{equation}
the Hubbard-Stratonovich field mediating dipole hopping
\begin{equation}
\label{eq:D_loc}
D_{\dot{\alpha}\mu}^{(0)} = - V_{\dot{\alpha}\mu}
\end{equation}

\begin{equation}
\label{eq:D_F}
\begin{split}
D^{F}_{\dot{\alpha} \mu}(t_1, t_2) =  & \ \frac{i}{2} \Pi^{F}_{\dot{\alpha} \alpha}(t_1, t_2) \delta_{\alpha \mu} +  \Big (  \int_{0}^{t_1}dt \, \Pi^{\rho}_{\dot{\alpha}\sigma }(t_1, t)V_{\sigma \dot{\beta}} D^{F}_{\dot{\beta} \mu} (t, t_2) \\
& - \int_{0}^{t_2}dt \, \Pi^{F}_{\dot{\alpha}\sigma }(t_1, t)V_{\sigma \dot{\beta}} D^{\rho}_{\dot{\beta} \mu} (t, t_2) \Big )
\end{split}
\end{equation}

\begin{equation}
\label{eq:D_P}
D^{\rho}_{\dot{\alpha} \mu}(t_1, t_2) =  \frac{i}{2} \Pi^{\rho}_{\dot{\alpha} \alpha }(t_1, t_2) \delta_{\alpha \mu} +  \int_{t_2}^{t_1}dt \, \Pi^{\rho}_{\dot{\alpha}\sigma }(t_1, t)V_{\sigma \dot{\beta}} D^{\rho}_{\dot{\beta} \mu} (t, t_2) \ ,
\end{equation}
and its corresponding self energies
\begin{equation}
\label{eq:Pi_F}
\Pi^{F}_{\dot{\sigma}\sigma}(t_1, t_2) = -2 \Big( F_{\sigma_{-}}(t_1, t_2) F^{*}_{\sigma_{+}}(t_1, t_2) - \frac{1}{4} \rho_{\sigma_{-}}(t_1, t_2) \rho^{*}_{\sigma_{+}}(t_1, t_2) \Big ) 
\end{equation}

\begin{equation}
\label{eq:Pi_P}
\Pi^{\rho}_{\dot{\sigma}\sigma}(t_1, t_2) = -2 \Big ( \rho_{\sigma_{-}}(t_1, t_2) F^{*}_{\sigma_{+}}(t_1, t_2) + F_{\sigma_{-}}(t_1, t_2) \rho^{*}_{\sigma_{+}}(t_1, t_2) \Big ) \ .
\end{equation}
When making use of the relations between left-dotted and right dotted quantities collected in the following section, these form a closed set of equations allowing for numerical time evolution.
We have made use  of the fact that $G, \Sigma, O, \Omega$ and $ \Pi$ are diagonal in spatial indices for all times when starting from initial states where this holds, as is the case for all initial states considered in this work.

\subsubsection{Green's Function Symmetries}
\label{app:symmetries}
We collect some symmetries between different matrix elements of Green's functions as well as consequences for their matrix elements following from the EOMs at NLO in this appendix. 
From their respective operator expression it follows immediately that 
\begin{equation}
F(t_2, t_1) = F(t_2, t_1)^{*} , \qquad \rho(t_2, t_1) = - \rho(t_1, t_2)^{*}
\end{equation}
and as a consequence via the KB equations

\begin{equation}
\Omega^{F}(t_2, t_1) = \Omega^{F}(t_1, t_2), \qquad \Omega^{\rho}(t_2, t_1)= -\Omega^{\rho}(t_1, t_2)^{*} \ .
\end{equation}
From their KB equations (\ref{eq:Om_F}, \ref{eq:Om_P}) we further observe $\Omega^{F}, \Omega^{\rho} \in \mathbb{R}$. One may use Eq. (\ref{eq:dGamma/dG}) to derive KB equations for $\Pi^{F}_{\sigma \dot{\sigma}}, \Pi^{\rho}_{\sigma \dot{\sigma}}$ in addition to those for $\Pi^{F}_{ \dot{\sigma}\sigma}, \Pi^{\rho}_{ \dot{\sigma} \sigma}$. However these turn out to be redundant and equivalent to the relations 

\begin{equation}
\Pi^{F / \rho}_{\sigma \dot{\sigma}} = \Pi^{F/ \rho \, * }_{\dot{\sigma} \sigma}\ .
\end{equation}
Making use of these equations it follows that 

\begin{equation}
\Pi^{F}(t_2, t_1) = \Pi^{F}(t_1, t_2), \qquad \Pi^{\rho}(t_2, t_1)= -\Pi^{\rho}(t_1, t_2)^{*}
\end{equation}
and then via induction over the KB equations that

\begin{equation}
D^{F / \rho}_{\sigma \dot{\sigma}} = -D^{F/ \rho \, * }_{\dot{\sigma} \sigma}
\end{equation}

\begin{equation}
D^{F}(t_2, t_1) = -D^{F}(t_1, t_2)^{*}, \qquad D^{\rho}(t_2, t_1)= D^{\rho}(t_1, t_2)^{*}
\end{equation}

\begin{equation}
O^{F}(t_2, t_1) = O^{F}(t_1, t_2), \qquad O^{\rho}(t_2, t_1)= -O^{\rho}(t_1, t_2) \ .
\end{equation}
For the family of initial states considered in this work it is also a notable consequence that $O^{F/ \rho}_{ij} \propto \delta_{ij}$ for all times as is simple to prove via induction by making use of the KB equations.

\section{Energy Density in 2PI From
Galitskii-Migdal Formula}
\label{app:Energy_density}
We compute different conserved densities of interest. Noting that the time derivative of the statistical 2-point function is given by 
\begin{equation}
\partial_{t_1} F_j (t_1, t_2)|_{t_{1} = t_{2}} = \frac{1}{2} \big \langle [H, a_{1}^\dagger] a_{1}   + a_{1} [H, a_{1}^\dagger ] \big \rangle 
\end{equation}
and making use of the commutation relations for creation and annihilation operators one obtains the relation
\begin{equation}
\partial_{t_1} F_j (t_1, t_2) |_{t_{1} = t_{2}} = i \big \langle h_{j-1}^D + 2 h_{j}^{D \dagger} + h_{j+1} + 2 h_{j}^{U} + 2 U n_{j} \big \rangle
\end{equation}
where $h_l^{D} = a_{l-1} a_{l}^{\dagger 2}a_{l+1}$ and $h_{j}^{U} = U n_j (n_j-1)$ represent the dipole and density-density interaction parts of the Hamiltonian
\begin{equation}
\label{eq:H_decomposition}
H = \sum_{j}  (h_j^{D} + h_j^{D \dagger})  + h_{j}^{U} \ .
\end{equation}
Combining these relations we observe the fact that 
\begin{equation}
E = \langle H \rangle = \frac{1}{2i} \sum_j \partial_{t_1} F_j (t_1, t_2)|_{t_{1} = t_{2}}- U\sum_j \langle n_j \rangle 
\end{equation}
allowing us to define a local energy density as
\begin{equation}
E_{j}(t_1) = \text{Re} \Big (\frac{1}{2i}  \partial_{t_1} F_j(t_1, t_2)|_{t_{1} = t_{2}} - U_{j} \big( F_{j}(t_1, t_1)  - \frac{1}{2}\big )\Big ) \ .
\end{equation}

\section{Preparation of Non-Gaussian Initial States}
We go beyond initial states with Gaussian density matrix by solving self-consistently the Kadanoff-Baym equations in equilibrium. This restriction to equilibrium is achieved by the fact that two-time functions like $F(t_1, t_2)$ and $\ O^{F/\rho}(t_1, t_2)$ are independent of the central time $T = (t_1 + t_2)/2$ and depend only on the relative time $\tau = t_1 - t_2$. Further, in equilibrium propagators fulfil their respective fluctuation dissipation relations, that is after Fourier transforming with respect to $\tau$ we have $F(\omega) = i \big (n_{\beta, \mu} (\omega) - \frac{1}{2} \big ) \rho(\omega)$
with $n_{\beta, \mu} (\omega) = 1/( e^{\beta(\omega - \mu)} - 1)$. The auxiliary correlators $D$ and $O$ fulfil identical relations with $\mu = 0$ since their corresponding operators carry no charge under the particle-number operator. We further restrict to $J = 0$ which corresponds to dropping the dipole-hopping mediating auxiliary field, i.e $D = 0$. Introducing retarded propagator components as $A^{R}(t) = A^{\rho}(t) \theta(t)$ the Kadanoff-Baym equations in equilibrium become
\begin{equation}
\label{eq:P_Rw}
\rho^{R}(\omega) =\frac{1}{2 \pi} \big (\omega + \Sigma^{(0)} +  2 \pi \Sigma^{R}(\omega) \big )^{-1} \delta_{ij}  
\end{equation}

\begin{equation}
\label{eq:O_Rw}
O^{R}(\omega) =  \big (1 -2 \pi U \Omega^{R}(\omega)\big )^{-1} i U \Omega^{R}(\omega) U
\end{equation}

\begin{equation}
\label{eq:O_Fw}
O^{F}(\omega) =  \big (1 -2 \pi U \Omega^{R}(\omega)\big )^{-1}  U \Omega^{F}(\omega) \big(i U + 2 \pi O^{R}(-\omega)\big)\ .
\end{equation}
We solve these equations self-consistently by iterative schemes. The resulting propagators are then inserted into a $t_1 - t_2$-plane subset $\{(t_1, t_2) | t_1, t_2 < T\}$, where $T$ denotes a time by which the relative-time propagators have sufficiently decayed, and taken as initial conditions for the real-time evolution. In the present work we prepare local thermal states under the on-site interacting part of the Hamiltonian $H^{U} = U n (n-1)$. The energy and charge profiles are tuned by varying the conjugate parameters $\beta, \mu$ which enter the solver through FDRs. The obtained product states have identical energy-profiles under both the full and restricted Hamiltonians. While this method does not allow us to directly prepare thermal states for $U = 0$, diffusion constants for vanishing $U$ are obtained by time evolving for a short time under Hamiltonian with $U = 1$ and quenching to $U =0$ after some potential energy has been converted to kinetic energy.
\label{app:selfConsistent}

\section{Short-Time Expansion}
\label{app:short_time}
 Here we derive the short-time expansion for the density profile of a Gaussian initial state with a local inhomogeneity.
In an uncorrelated Gaussian initial state $\rho = \bigotimes_{i} \rho_{i}$ the short-time dynamics of charges can be calculated explicitly. We assume that the considered state exhibits a homogeneous profile of the charge expectation value with a single inhomogeneity localized on one lattice site, i.e. $\langle \hat{n}_{i}(t=0) \rangle = n_{B} + \delta_{i0} \delta n$, as considered in the main text.

In general, the short time expansion for the charge expectation values is given by
\begin{equation}
\label{eq:short_time_charge_expansion}
    \langle \hat{n}_{i}(t) \rangle = \Tr \rho \hat{n}_{i} + i t \, \Tr \rho [H, \hat{n}_{i}] + \frac{(it)^2}{2} \Tr \rho [H, [H, \hat{n}_{i}]] + \dots \ .
\end{equation}
Using the decompositon of the Hamiltonian \eqref{eq:H_decomposition} some algebra reveals the relations
\begin{equation}
    \sum_{i} [h_{i}^{D}, n_{j}] = (\Delta h^{D})_{j} \equiv h_{j+1}^{D} - 2 h_{j}^{D} + h_{j-1}^{D} , \qquad \sum_{i} [h_{i}^{D \dagger}, n_{j}] = -(\Delta h^{D \dagger})_{j} \ .
\end{equation}
Since $[h^{U}_{i}, \hat{n}_{j}] = 0$ and expectation values of $h^{D}$ and $h^{D \dagger}$ vanish in product states the linear term in the above expansion \eqref{eq:short_time_charge_expansion} must vanish. The relevant commutator for the quadratic term is then
\begin{equation}
    [h^{D}_{i}, h_{i}^{D \dagger}] = \hat{n}_{i}(\hat{n}_{i}-1) (\hat{n}_{i+1} + 1) + \hat{n}_{i-1} (\hat{n}_{i} + 1) (\hat{n}_{i} + 2) - 2 \hat{n}_{i-1} (2 \hat{n}_{i} + 1)(\hat{n}_{i+1} + 1)
\end{equation}
since expectation values of the same commutator on different sites vanish, i.e. $i \neq j \implies [h^{D}_{i}, h_{j}^{D \dagger}] = 0$. Making use of Wick's theorem, which for Gaussian product states implies $\langle \hat{n}_{i}^2 \rangle = 2 \langle \hat{n}_{i} \rangle^2 + \langle \hat{n}_{i}\rangle$, as well as the density profile of the initial state we obtain the short-time expansion for the charge dynamics on the center site as 
\begin{equation}
    \langle \hat{n}_{0}(t) \rangle = n_{B} + \delta n - t^{2} \times 2\Big( 6 \delta n \, n_{B}^2 + n_{B} (4 \delta n^2 + 6 \delta n) + 2 \delta n^2 \Big) + \mathcal{O}(t^3) \ .
\end{equation}
This expansion agrees with simulation results due to the fact that all Feynman diagrams contributing orders up to $t^2$ have been included in the 2PI approximation. A comparison is shown in Fig. \ref{fig:ScalingN}.

\end{widetext}
\bibliography{bibliography.bib}
\end{document}